# Characterization of Al–W oxide coatings on aluminum formed by pulsed direct current plasma electrolytic oxidation at ultra-low duty cycles

Kristina Mojsilović[1], Nenad Tadić[1], Uroš Lačnjevac[2], Stevan Stojadinović[1], Rastko Vasilić[1]*

[1] *University of Belgrade, Faculty of Physics, Studentski trg 12-16, 11000 Belgrade, Serbia*

[2] *University of Belgrade, Institute for Multidisciplinary Research, Kneza Višeslava 1, 11030 Belgrade, Serbia*

*Corresponding author. Tel: + 381-11-7158161; Fax: + 381-11-3282619

E-mail address: Rastko.vasilic@ff.bg.ac.rs (Rastko Vasilić)

## Abstract

The growth of thin oxide coatings on the aluminum substrate in water-based sodium tungstate electrolyte by plasma electrolytic oxidation (PEO) is discussed and experimentally illustrated. The growth is carried out using a distinctive ultra-low duty cycle pulsed direct current (DC) power supply. During the PEO processing elements present in micro-discharges are identified using standard optical emission spectroscopy (OES) technique. The spectral line shape analysis of the first two hydrogen Balmer lines shows the presence of two types of micro-discharges. Obtained coatings are also characterized with respect to their morphology and chemical and phase composition. It is shown that coatings are composed of Al, O, and W, featuring low roughness and porosity. Partial crystallization of the coatings resulted in identification of $WO_3$, $W_3O_8$, and $\gamma$-$Al_2O_3$ crystalline phases.

## 1. Introduction

Plasma electrolytic oxidation (PEO) is an efficient and environmentally friendly surface engineering process used on various metals (Al, Mg, Ta, Ti, Zr, etc.) and their alloys to create oxide coatings [1]. PEO process differs from conventional anodizing by using high voltages (greater than the dielectric breakdown voltage of the original oxide film) and diluted alkaline electrolytes. This leads to the formation of plasma as indicated by the appearance of numerous micro-discharges on the metal surface, accompanied by gas evolution and locally increased temperature and pressure. Due to these new thermal and chemical conditions, the structure, morphology and composition of the growing oxide coatings change. The oxide coatings obtained by this process contain elements from both substrate and electrolyte solution, and have good electrical and thermal properties, strong adhesion, high corrosion and wear resistance. Those characteristics can be modified by applying different electrical regime with DC, pulsed-DC or AC power source, thus varying micro-discharge intensity in order to obtain oxide coatings of desired properties [2].

Various research groups have shown that using a pulsed DC power source can provide better control over plasma micro-discharges than a continuous DC power source, and with that, improved homogeneity, higher microhardness, reduced thickness of the porous layer and higher corrosion resistance [3-5]. This can be achieved by varying electrical parameters such as voltage and current density, as well as by setting the duration of pulses − $t_{on}$, and time between pulses − $t_{off}$, which allows the system to return to equilibrium.

Although PEO processing of aluminum and its alloys is commonly performed in silicate containing electrolytes [6-8], it was shown that an addition of sodium tungstate to electrolyte results in the decrease of the breakdown voltage and increase of coatings' thickness, density and corrosion resistance [9-12]. It was also observed that the utilization of low concentrated sodium tungstate electrolyte produces thin PEO coatings [13], while the PEO of aluminum in an electrolyte with high sodium tungstate concentration yields the formation of micro- and nano-sized $WO_3$ precipitate in the solution [14].

In the present paper, a unique, ultra-low duty cycle pulsed DC electrical regime was applied in studying the PEO process on high purity Al in sodium tungstate solution. Used power supply features constant 2.5 A current controlled pulses with $t_{on}$ time ranging from 1 to 20 ms and much longer $t_{off}$ time which can be set from 1 to 2 s. If one takes into account that the radiation intensity during the PEO is rather low, i.e., long exposure times are required. Prolonged $t_{off}$ time (coupled with appropriate exposure) enables a recording of the radiation emitted during just one pulse by the detection system.

The main goal of the presented research was to investigate the utilization of this power supply for PEO processing and to characterize obtained oxide coatings. Optical emission spectroscopy (OES) was employed for in-situ characterization of plasma micro-discharges and determination of electron number density $N_e$. Investigation of surface morphology, chemical and phase composition of obtained oxide coatings was conducted using scanning electron microscopy (SEM/EDS) and X-ray diffraction (XRD).

## 2. Experimental

Rectangular samples of 1050 grade aluminum alloy were used as the anode material, and two platinum wires were used as cathodes in experiments. The anode material was sealed with insulation resin leaving an active surface area of approx. 2.5 $cm^2$ accessible to electrolyte. PEO process was carried out in an electrolytic cell with flat glass windows. Water solution of 0.01 M sodium tungstate ($Na_2WO_4 \cdot 2H_2O$) was prepared for all experiments in order to prevent the appearance of $WO_3$ precipitate in the solution [13, 14]. Electrolyte was prepared using double distilled and deionized water and p.a. (pro analysis) grade chemical compounds. During PEO, electrolyte circulated through chamber-reservoir system and the temperature of the electrolyte was kept at (22$\pm$2) °C. A home-made pulsed DC power supply working in galvanostatic (current controlled) mode was used for this experiment. The power supply produced rectangular pulses of 2.5 A with a possible $t_{on}$ value ranging from 1 to 20 ms and a $t_{off}$ value from 1 to 2 s. PEO processing was performed under a current density of 1 A/cm$^2$ during 10 min, 20 min and 30 min with a $t_{on}$ of 5 ms, 10 ms and 20 ms, and $t_{off}$ of 1 s for all experiments (Figure 1). Voltage pulse sequences were recorded using Tektronix TDS 2022 digital storage oscilloscope and a high voltage probe which was connected directly to the power supply (a current of 2.5 A was used for all experiments – this current cannot be changed).

For the indicative presence of spectral lines and an overview of the optical spectrum in wavelength range 360-850 nm, a low-resolution spectrometer was used (Ocean Optics), with integration time set to 10 s.

H$_\alpha$ and H$_\beta$ lines were recorded using 0.3 m Rank Hilger spectrometer (diffraction grating 1200 grooves/mm and inverse linear dispersion of 2.7 nm/mm) equipped with thermoelectrically cooled ICCD (-20 °C), with integration time of 1 s and instrumental profile with full width at half maximum (FWHM) of (0.23 $\pm$ 0.02) nm. For these high-resolution experiments, integration time of 1 s was selected because it enables the recording of optical emission spectra originating from just one $t_{on}$ pulse.

A scanning electron microscope (SEM, JEOL 840A) was used to examine surface morphology of the PEO layers, and chemical composition of their surface was analyzed with energy dispersive spectrometer (EDS) coupled with SEM. For cross-sectional SEM analyses, samples were embedded in epoxy resin and polished with 220, 1000, and 4000 SiC abrasive papers, followed by polishing with 1 μm diamond paste. Cross-sectional micrographs were obtained acquired on Tescan VEGA TS 5130 MM with BSE detector. Rigaku Ultima IV diffractometer with Ni-filtered CuK$_\alpha$ radiation source was used for crystal phase identification. Crystallographic data was collected in Bragg-Brentano mode, in $2\theta$ range from 10° to 90° with a scanning rate of 2 °/min. Porosity and surface roughness of obtained coatings were estimated using ImageJ software.

## 3. Results and discussion

### 3.1. Spectral characterization of the PEO process

Optical emission spectra of micro-discharges during PEO of aluminum in sodium tungstate solution in the wavelength range 360-850 nm are presented in Figs. 2 and 3. The main difficulty in OES characterization is embodied by the space and time inhomogeneity of micro-discharges appearing randomly across the anode surface. The species that are identified originate either from the aluminum electrode or from the electrolyte, and they were identified using the NIST (National Institute for Standards and Technology) online spectral database [15]. The registered lines, in the decreasing intensity order, are: Al I at 394.40 nm and 396.15 nm, $H_\alpha$ at 656.28 nm, $H_\beta$ at 486.13 nm, Na I at 588.99 nm and 589.59 nm, and O I at 777.19 nm, 777.42 nm and 777.54 nm, where notation I refers to neutral atoms. Apart from atomic lines, a strong AlO molecular band was also detected. The registered continuum emission comes from a collision-radiative recombination of electrons (bound-free radiation) [16] and the bremsstrahlung radiation (free-free radiation) [17] emitted from the growing oxide coating. The collision-radiative recombination of electrons results in light emission or luminescence during recombination, while bremsstrahlung is the electromagnetic radiation produced by the deceleration of charged particles, which is caused by the deflection by other charged particles. Therefore, observed spectra are a combination of the radiation originating from the micro-discharges and the radiation from the oxide coating itself.

The analysis of the optical emission spectra recorded at 10 min, 20 min and 30 min of PEO with $t_{on}$ varying from 5 ms to 20 ms shows that PEO spectral line intensities with $t_{on} = 5$ ms are lower than with $t_{on}$ being 10 ms or 20 ms. It can also be noticed that the intensity of the plasma micro-discharging radiation shows lower values at 30 min than at 20 min from the beginning of the experimental process (see Fig. 3), since the percentage of the oxide coating area simultaneously covered by active discharge sites decreases during the PEO process [1]. It can also be observed from Fig. 3 that spectra acquired at lower PEO processing time are dominated by the Al I spectral line resulting in bluish appearance of the micro-discharges, while with prolonged processing the Na I spectral line becomes stronger thus coloring micro-discharges more orange [18].

The electron number density calculations are useful for studying plasma processes because they give us information how many free electrons there are in a given volume. Since all plasmas have some degree of ionization, this means that there are electrons that have been stripped from atoms, and are moving under the applied electric field [19]. To determine the PEO electron number density, $N_e$, the plasma broadened profiles of the $H_\alpha$ and the $H_\beta$ lines were used. During the analysis of the $H_\alpha$ line profile (Fig. 4), it was found that the $H_\alpha$ line shape can be properly fitted only with the use of two Voigt profiles, which is the consequence of the used spectrometer having large instrumental profile of a Gaussian shape (FWHM= $(0.23 \pm 0.02)$ nm). These Voigt profiles, according to Eq. (10) in [20] correspond to electron number densities represented in Table 1 and Table 2, depending on duration of the PEO process and $t_{on}$ value. The values represented as $N_{e1}$ result from FWHM values of narrow and those represented as $N_{e2}$ from FWHM values of wide profiles.

However, results coming from the $H_\alpha$ line analysis are not considered confident, since self-absorption of such strong line can broaden the line considerably while the line shape can still be fitted with two Voigt profiles. To avoid the interference of the AlO molecular band with the $H_\beta$ line on aluminum (see Fig. 5), the same line was recorded from PEO with tantalum anode where AlO molecular bands are not present. This procedure was already conducted in the literature [21] under the assumption that the characteristics of dielectric breakdown through the oxide layer differ depending on anodic materials. Namely, for high melting point anodic materials (such as Ti and Ta) dielectric breakdown of the oxide layer occurs, but melting and ejection of the anode material does not occur, i.e., there is no interference of anode material containing spectral lines (AlO in the present study) with the $H_\beta$ line [22]. Furthermore, $N_e$ values obtained for various substrate materials (Table 2 in Ref. 22) are quite similar, suggesting that this procedure may be applied. During the analysis of the $H_\beta$ line profile (Fig. 6), it was found that the $H_\beta$ line shape can also be properly fitted only with the use of two Voigt profiles. These Voigt profiles, according to Eq. (2a) in [23] correspond to electron number densities of $N_{e1} = 0.85 \cdot 10^{15}$ cm$^{-3}$ and $N_{e2} = 2.45 \cdot 10^{16}$ cm$^{-3}$, that result from FWHM values of narrow and from FWHM values of wide profiles, respectively. The calculated values of electron number density from $H_\beta$ line analysis prove the presence of self-absorption in the $H_\alpha$ line. These results are in agreement with previously reported results for PEO processing of tantalum in direct current mode [22, 24], proposing that micro-discharges in pulsed-DC and DC mode are characterized by the same parameters.

The measured values of electron number density that result from $H_\alpha$ and $H_\beta$ line analysis, could be summarized using the PEO model presented in [25]. According to the mentioned model and optical emission spectroscopy study on PEO of aluminum [21], the low and medium values of electron number densities would correspond to A and C type discharges (see Fig. 9 in [25]), which are the discharges happening in the micropores at the surface of oxide coating or near the surface of oxide coating. The highest values of electron number density could relate to type B discharge when the molten aluminum can be ejected from the substrate directly to the electrolyte, but the proposed model cannot explain the formation of micro-discharges through dielectric in the case of high melting temperature anodes like Ti and Ta when no ejection of the anode material occurs.

Low $N_e$ values are usually related to the breakdown plasma occurring in electrolyte at the surface of the dielectric, while medium $N_e$ values can be related to the breakdown through the dielectric oxide layer further from the electrolyte/oxide surface towards anode [22]. Consequently, high $N_e$ values originate from the third process occurring after the breakdown of dielectric and it is followed by ejection of anodic material plasma towards the surface. For high melting point materials (such as Ti and Ta) only breakdown of the dielectric occurs, while melting and the ejection of the anode material does not occur. Clearly, $N_e$ values (representing the ionization level of plasma) are strongly related to the region where the discharging process takes place. In order to have further insight into the observed $N_e$ values and processes occurring during the PEO, timer resolved spectroscopy of a single PEO micro-discharge has yet to be performed.

The observed optical emission spectra are space-time integrated radiation recorded by a spectrometer-detector system. In this study spectra are time averaged over 1 s, while space averaging has been done over thousands of short living micro-discharges distributed over the surface of the sample. Therefore, the number of discharging processes may be higher than that proposed in [25]. For example, Cheng et al. [26] proposed a modified PEO discharging model which includes two additional discharge types taking place inside of the oxide coating, making them hardly visible for OES.

### 3.2. Morphology, chemical, and phase composition of PEO coatings

In order to determine the surface morphology and chemical composition of the coatings formed by PEO of Al, a SEM/EDS analysis was performed. Fig. 7 shows surface morphology evolution of the PEO coatings obtained for applied processing parameters, while Fig. 8 shows the cross-sections of the coatings formed under the same conditions.

The surface of the oxide coatings is decorated with pores of different shapes and sizes, and molten regions that appear as a result of the rapid cooling of the electrolyte, which is inherent to the PEO process itself. The surface morphology of the obtained coatings is similar to the one reported by other groups in the same system, with slightly more pronounced porosity [27-29]. The most probable reason for the increased porosity comes from the fact that during the PEO processing instantaneous peak voltage pulses were as high as 1300 V (Fig. 1), as a result of unusually high current density applied for these experiments (1 A/cm$^2$) and low concentration of Na$_2$WO$_4$ containing electrolyte. Utilized ultra-low duty cycle value of about 1% allowed application of high voltages [6] and comprehensive cooling of the coating between the pulses, thus promoting the complete solidification of the molten material ejected from individual micro-discharges.

At the beginning of dielectric breakdown, the total current through the coating is a sum of currents through thousands of small (with respect to surface area) micro-discharging channels randomly distributed over the sample's surface. With prolonged PEO processing time, the coating thickens thus requiring higher current for each dielectric breakdown and the number of simultaneously active micro-discharging channels decreases. As a result, current is localized at weak points (flaws) in the oxide layer, and works its way through the oxide coating, but leaves behind a discharge channel of a larger diameter. In other words, the number of the micro-discharging channels decreases, but the current inside of each micro-discharging channel increases in order to keep the total value of current unchanged. Consequently, the size of the pores which are observable on the coating increases, but their spatial density and overall porosity decrease. Our experimental data show a decrease in porosity from 7.5 % to 5.0 % with the increasing $t_{on}$ and processing time (Table 3). Encouraged by the porosity estimation results (typical PEO coatings have overall porosity of about 10 - 20 % [2]), we performed an estimate of the surface roughness of the obtained coatings (Table 3). The obtained coatings have unusually low surface roughness ($R_a$), which is about one order of magnitude lower than surface roughness estimates for similar systems [28-

]. On the other hand, it should be noted that surface roughness increases with the thickness of oxide coatings, so the discrepancy in the surface roughness may be related to coatings' thickness (coatings obtained in referenced studies are in average 20 - 40 µm thick, compared to 1 - 3 µm in this paper).

In order to determine the chemical composition of obtained PEO coatings, we performed a set of EDS analyses (Table 4). The main elements of the coatings are Al, O, and W. The EDS analysis shows an increase in concentration of O and W, and a decrease in concentration of Al during time and with an increasing $t_{on}$ value. This is highly expected since at the beginning of the PEO process barrier layer is formed which mainly consists of the substrate material oxide. After the breakdown, electrolyte species, drawn by the high electric field are being incorporated into the oxide coating and their content increases with prolonged PEO times [1].

EDS analysis was also performed at characteristic points of the surface coating. Figure 9 shows a sample processed for 20 min with $t_{on}$ set to 20 ms, while Table 5 contains the results of this analysis. Spectrum 1 represents the analysis of the whole surface of the coating, Spectrum 2 is the analysis in the micro-discharge channel, and Spectrum 3 is the analysis of the most prominent part of the oxide coating. Clearly, the concentration of Al is the highest inside of the micro-discharge channel due to oxidation of alumina during PEO process at the metal/oxide interface, while the concentration of W is the highest on the top surface of the coating. However, EDS elemental composition results should be taken with precaution, because the accuracy of elemental composition determination varies for elements under investigation (Al, W, and O) from 0.1 % for W to 8 % for O (given as standard deviation by the instrument).

The cross-sections of the oxide coatings (Figure 8) show an increase in PEO layer thickness with PEO processing time. The thickest oxide coating is around 3 µm thick and it is formed for $t_{on}$ = 20 ms during 30 min of PEO processing. Regardless of the processing time, the obtained coatings are very thin and non-continuous, which limits their possible applications. We believe that the observed discontinuity in the coatings comes as a result of the high voltages registered during the PEO processing which favor the destruction instead of accelerating the growth rate of the coating. In fact, the quality of the coatings is strongly related to the PEO processing conditions. High values of $t_{off}$ time (1 s) result in a large pause between the pulses, thus requiring higher processing times in order to grow thicker coatings. As previously reported, high current pulses are detrimental for the quality of PEO coatings, because longer PEO processing times result in fragmented coating coverage of the substrate surface [31].

XRD patterns of oxide coatings obtained after various PEO times and $t_{on}$ values are presented in Figure 10. For the sample processed with a $t_{on}$ time of 5 ms, one can observe only strong diffraction maxima originating from the substrate and rather low maxima originating from γ-$Al_2O_3$ and $WO_3$ phases. With prolonged PEO processing time, the $W_3O_8$ crystalline phase slowly appears and the diffraction maxima originating from Al substrate decrease as a result of increasing coating thickness. When $t_{on}$ times of 10 and 20 ms are applied, XRD patterns feature stronger appearance of the $W_3O_8$ phase, while γ-$Al_2O_3$ and $WO_3$ diffraction maxima become more pronounced as a result of a higher level of crystallization. Tungsten oxides are formed after the $WO_4^{2-}$ anions are generated (from ionization of sodium tungstate in water) and drawn into the discharge channels by means of a strong electric field [24]. At the same time, alumina is molten and ejected from the micro-discharge channels and it rapidly solidifies in the contact with the surrounding low temperature electrolyte at the coating/electrolyte interface.

It is also interesting to mention that we were not able to detect maxima corresponding to α-$Al_2O_3$, which commonly appear in this system (for example, see Ref [27]). Although it can be postulated that the reason for the absence of the $\alpha$-$Al_2O_3$ phase is related to the ultra-low duty cycle pulsed-DC regime and/or electrolyte we used in this study, it is quite unexpected that the transition from $\gamma$-$Al_2O_3$ to $\alpha$-$Al_2O_3$ induced by locally high temperatures [32] did not occur. Previously published results by Martin et al. [33] and Yerokhin et al. [34] reported post-discharge cooling rates of the oxide layer of about $10^7$ K/s, suggesting that the $t_{off}$ value of 1 s is much longer than required for successful cooling of the oxide layer. On the other hand, in both papers the $\alpha$-alumina phase is observed in the coatings, suggesting that further work is required to investigate this finding in details.

## 4. Conclusions

In the present study we investigated PEO on high purity Al in a sodium tungstate electrolyte using ultra-low duty pulsed-DC power source. This enabled us to perform OES and study the micro-discharge characteristics from one voltage pulse. We were able to identify species of the coatings originating from both the electrolyte and the substrate. Using the hydrogen Balmer $H_\alpha$ and $H_\beta$ lines, we calculated the electron number density of plasma during the PEO processing.

The oxide coating morphology changed with the PEO time and $t_{on}$ value. For lower $t_{on}$ values, coatings are fragmented as a result of high pulse current and/or voltage. Experimental data show a decrease in porosity from 7.5 % to 5.0 % with increasing the $t_{on}$ value and processing time, low surface roughness ($R_a \leq 1$ $\mu$m), and low thickness (1-3 $\mu$m) for all coatings. Continuous coatings with low surface porosity obtained with 20 ms pulses may be further investigated for corrosion resistance. EDS analysis confirmed the presence of Al, O, and W in the oxide coating. XRD analyses confirmed the gradual crystallization of the oxide coatings for all PEO processing times and the absence of the $\alpha$-$Al_2O_3$ crystalline phase, which may be related to the specific ultra-low duty pulsed DC source and/or electrolyte composition, but it requires further analyzing.

**Acknowledgements**

This work is supported by the Ministry of Education, Science, and Technological Development of the Republic of Serbia and by the European Union Horizon 2020 research and innovation program under the Marie Sklodowska-Curie grant agreement No. 823942.

**Figure captions:**

**Figure 1.** Waveforms used in experiments: a) pulse sequence with $t_{off}$ = 1 s; b) 5 ms pulse at 10 and 30 min of PEO processing; c) 10 ms pulse at 10 and 30 min of PEO processing; d) 20 ms pulse at 10 and 30 min of PEO processing.

**Figure 2.** Optical emission spectra recorded at 10 min of PEO process.

**Figure 3.** Optical emission spectra recorded with $t_{on}$ of 10 ms.

**Figure 4.** $H_\alpha$ spectral line during the PEO on aluminum (10 ms, 20 min): a) $H_\alpha$ line experimental profile fitted with two Voigt profiles; b) residue plot.

**Figure 5.** $H_\beta$ spectral line during the PEO of aluminum (10 ms, 20 min): The $H_\beta$ line experimental profile at 20 min from the beginning of the process with $t_{on}$ = 10 ms.

**Figure 6.** $H_\beta$ spectral line during the PEO of tantalum (10 ms, 20 min): a) $H_\beta$ line experimental profile fitted with two Voigt profiles; b) residue plot.

**Figure 7.** SEM micrographs of oxide coatings formed under applied PEO processing conditions.

**Figure 8.** SEM micrographs of oxide coatings cross sections formed under applied PEO processing conditions.

**Figure 9.** SEM micrograph of an oxide coating at $t_{on}$ = 20 ms and at 20 min of PEO time.

**Figure 10.** XRD patterns of oxide coatings formed at different PEO times with $t_{on}$ set to: a) 5 ms; b) 10 ms; c) 20 ms; identified crystalline phases: (1) Al; (2) $WO_3$; (3) $W_3O_8$; (4) $\gamma$-$Al_2O_3$.

**Table captions:**

**Table 1.** Electron number density values depending on the $t_{on}$ value (estimated error ±35%).

**Table 2.** Electron number density values depending on the duration of the experimental process (estimated error ±35%).

**Table 3.** Estimated surface roughness and porosity data for various PEO processing parameters.

**Table 4.** EDS analysis of the surface coatings depending on the $t_{on}$ value and duration of the experimental process.

**Table 5.** EDS analysis of surface coating at characteristic points.

Table 1

| $t_{on}$ [ms] | $N_{e1}[10^{16}\,\mathrm{cm^{-3}}]$ | $N_{e2}[10^{17}\,\mathrm{cm^{-3}}]$ |
|---|---|---|
| 5 | 1.0 | 5.0 |
| 10 | 0.7 | 4.4 |
| 20 | 1.2 | 5.4 |

Table 2

| PEO process duration [min] | $N_{e1}[10^{16}\,\text{cm}^{-3}]$ | $N_{e2}[10^{17}\,\text{cm}^{-3}]$ |
|:---:|:---:|:---:|
| 10 | 1.0 | 4.3 |
| 20 | 0.9 | 5.5 |
| 30 | 0.9 | 5.0 |

Table 3

| t₁ [ms] | PEO time [min] | Rₐ [μm] | Porosity (%) |
|---|---|---|---|
| 5 | 10 | 0.27 | n/a |
| | 20 | 0.34 | n/a |
| | 30 | 0.65 | n/a |
| 10 | 10 | 0.22 | 7.48 |
| | 20 | 0.24 | 7.03 |
| | 30 | 0.46 | 6.88 |
| 20 | 10 | 0.44 | 5.66 |
| | 20 | 0.53 | 5.28 |
| | 30 | 0.62 | 4.96 |

| Position | Atomic fraction (%) | | |
|---|---|---|---|
| | Al | O | W |
| Spectrum 1 | 31.39 | 64.51 | 4.10 |
| Spectrum 2 | 69.04 | 28.47 | 2.49 |
| Spectrum 3 | 17.12 | 76.55 | 6.33 |

Table 5

| Element | Atomic fraction (%) | | | | | | | | |
|---|---|---|---|---|---|---|---|---|---|
| | 5ms | | | 10ms | | | 20ms | | |
| | 10 min | 20 min | 30 min | 10 min | 20 min | 30 min | 10 min | 20 min | 30min |
| O | 44.32 | 53.28 | 51.87 | 49.26 | 54.59 | 52.30 | 52.68 | 64.51 | 71.67 |
| Al | 55.06 | 45.39 | 46.92 | 50.00 | 43.99 | 46.51 | 46.06 | 31.39 | 21.98 |
| W | 0.61 | 1.32 | 1.21 | 0.75 | 1.42 | 1.19 | 1.26 | 4.10 | 6.35 |

Figure 1

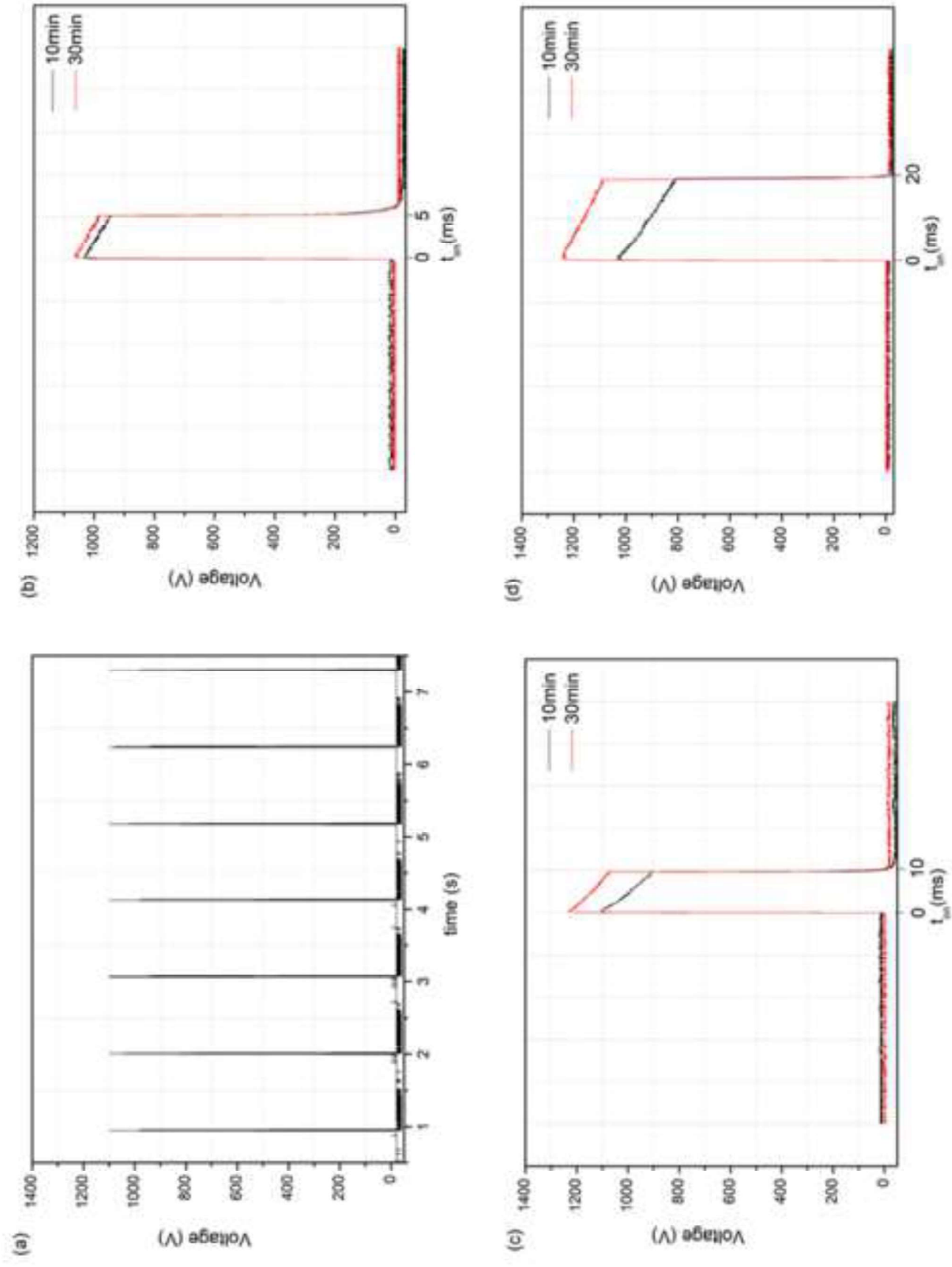

Figure 2

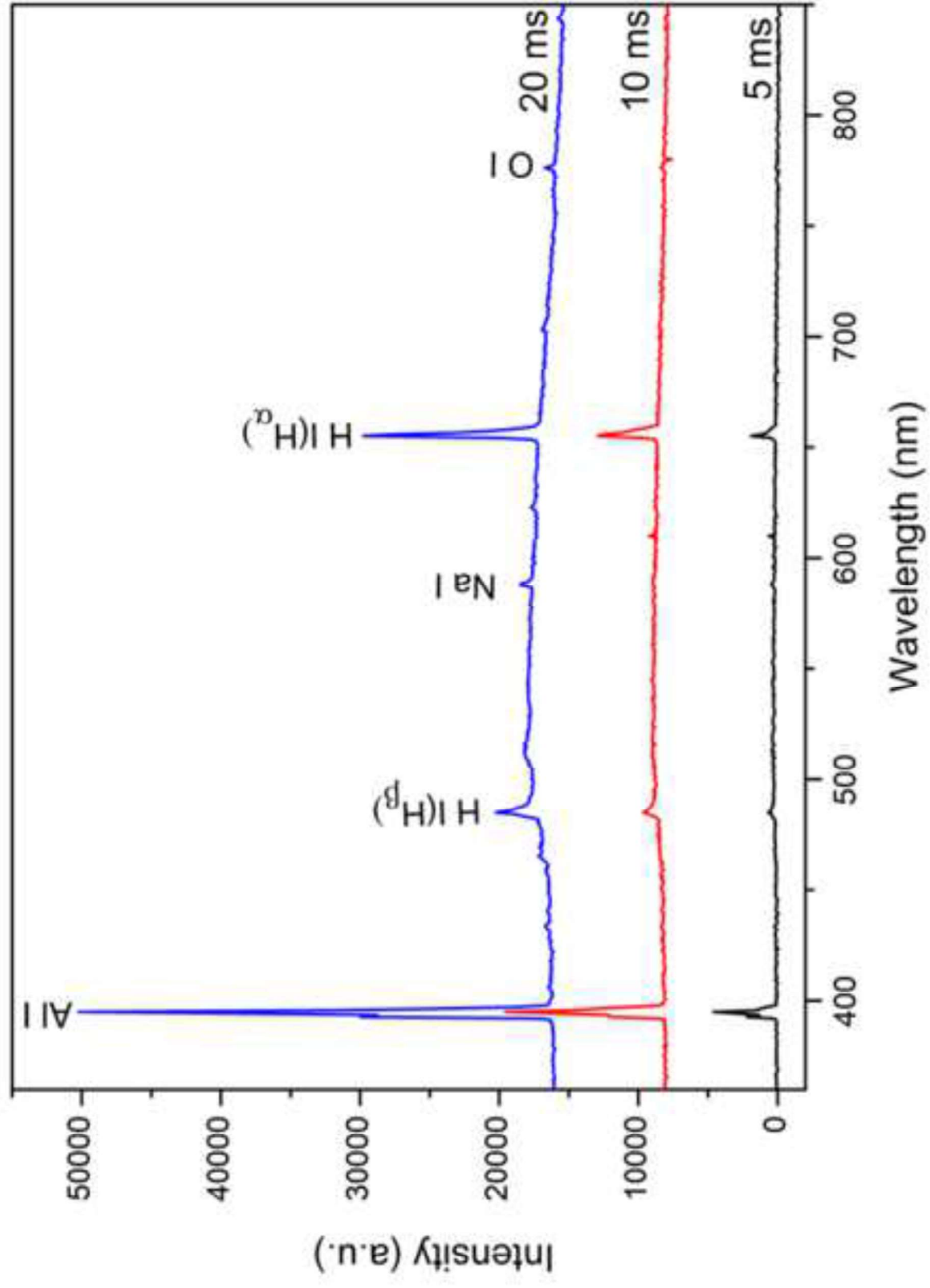

Figure 3

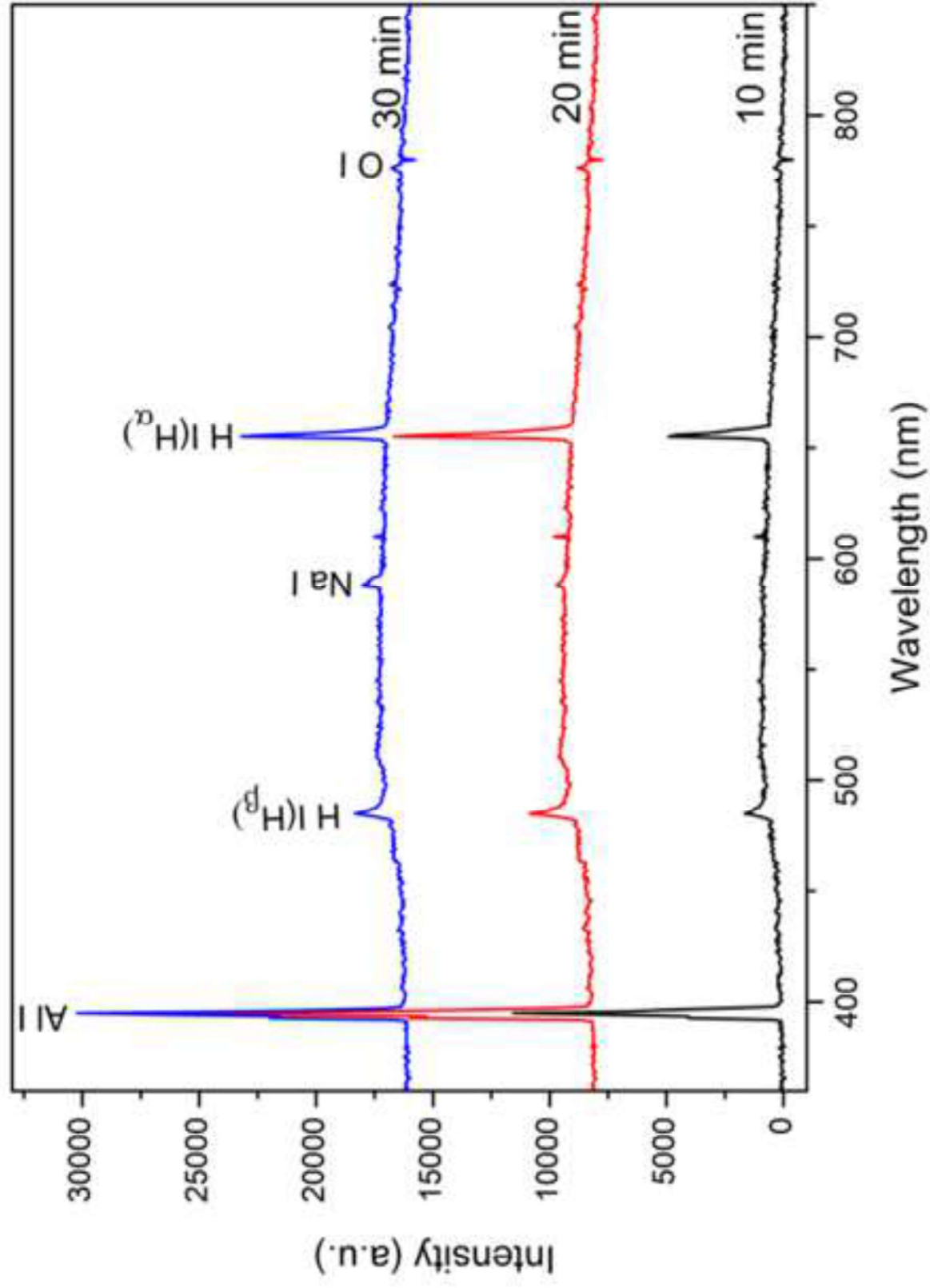

Figure 4

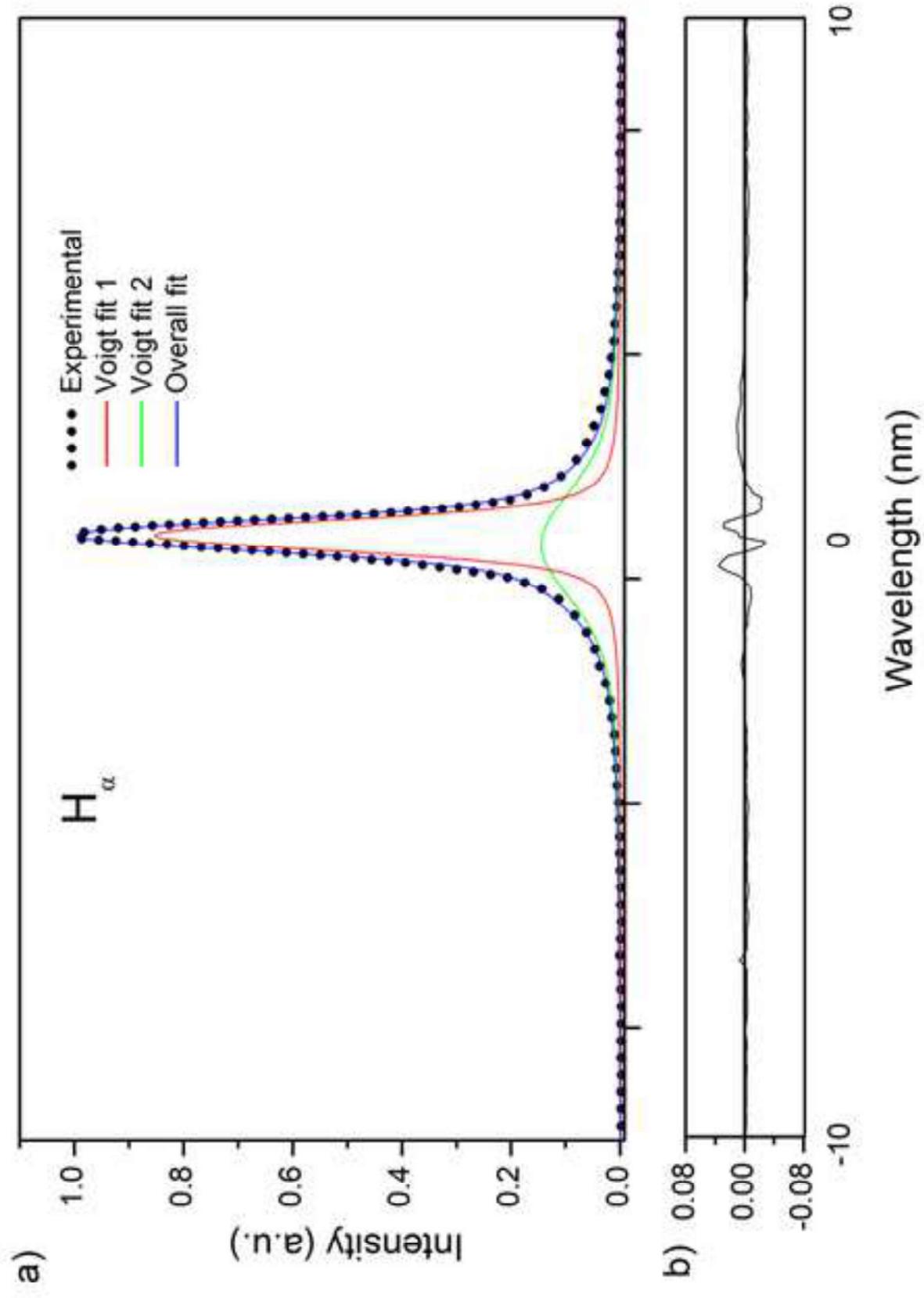

a) $H_\alpha$

Experimental
Voigt fit 1
Voigt fit 2
Overall fit

Intensity (a.u.)

b)

Wavelength (nm)

Figure 5

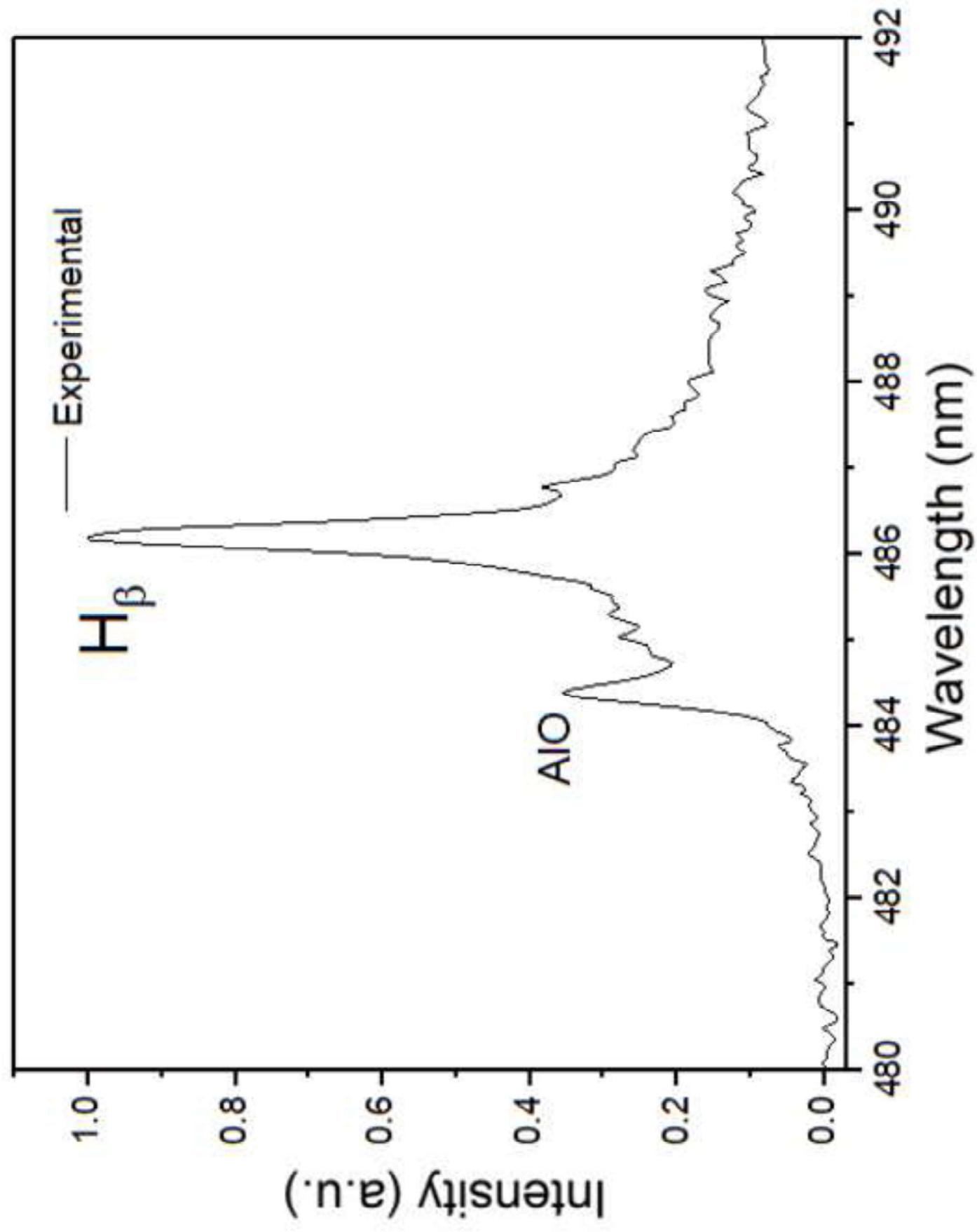

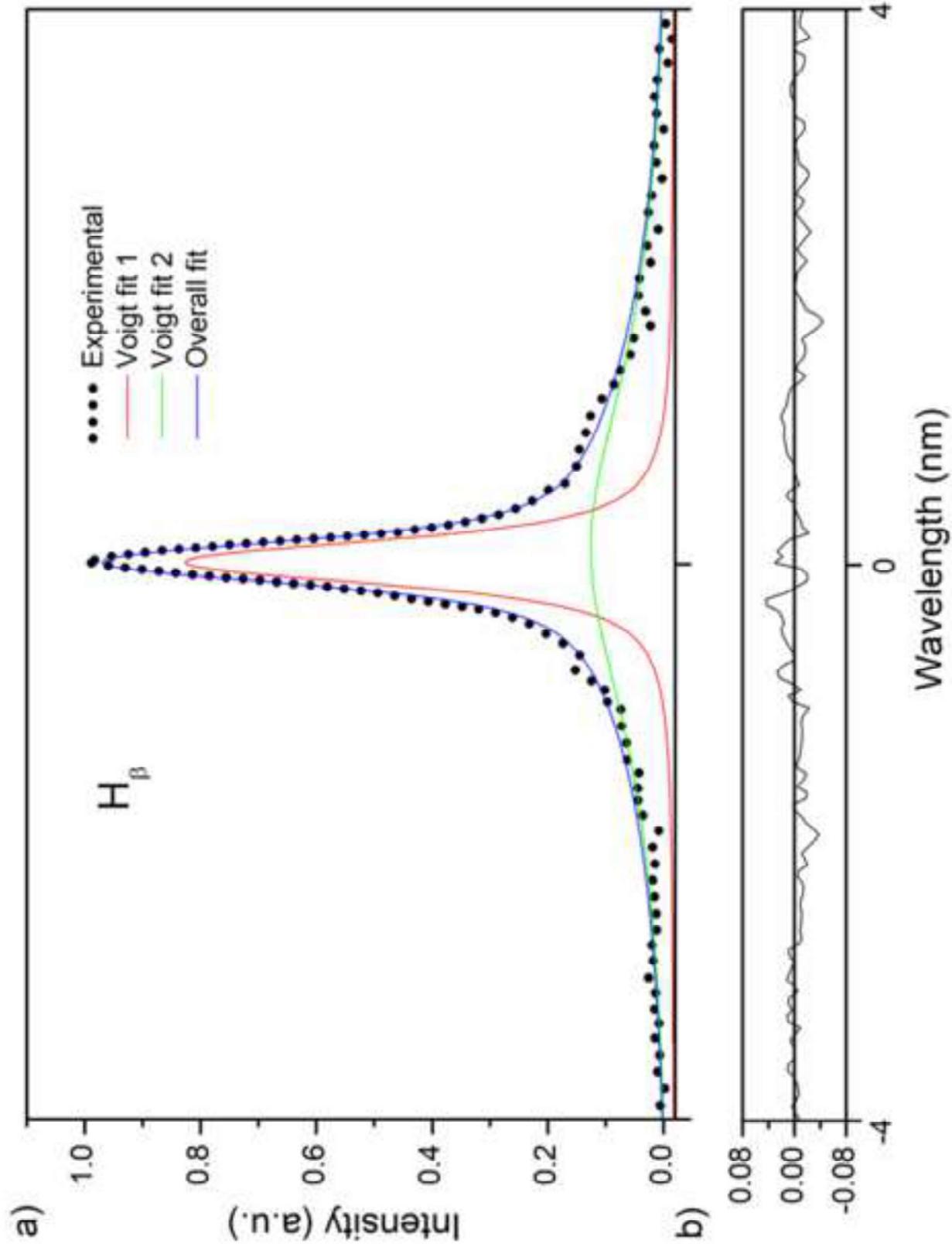

Figure 7

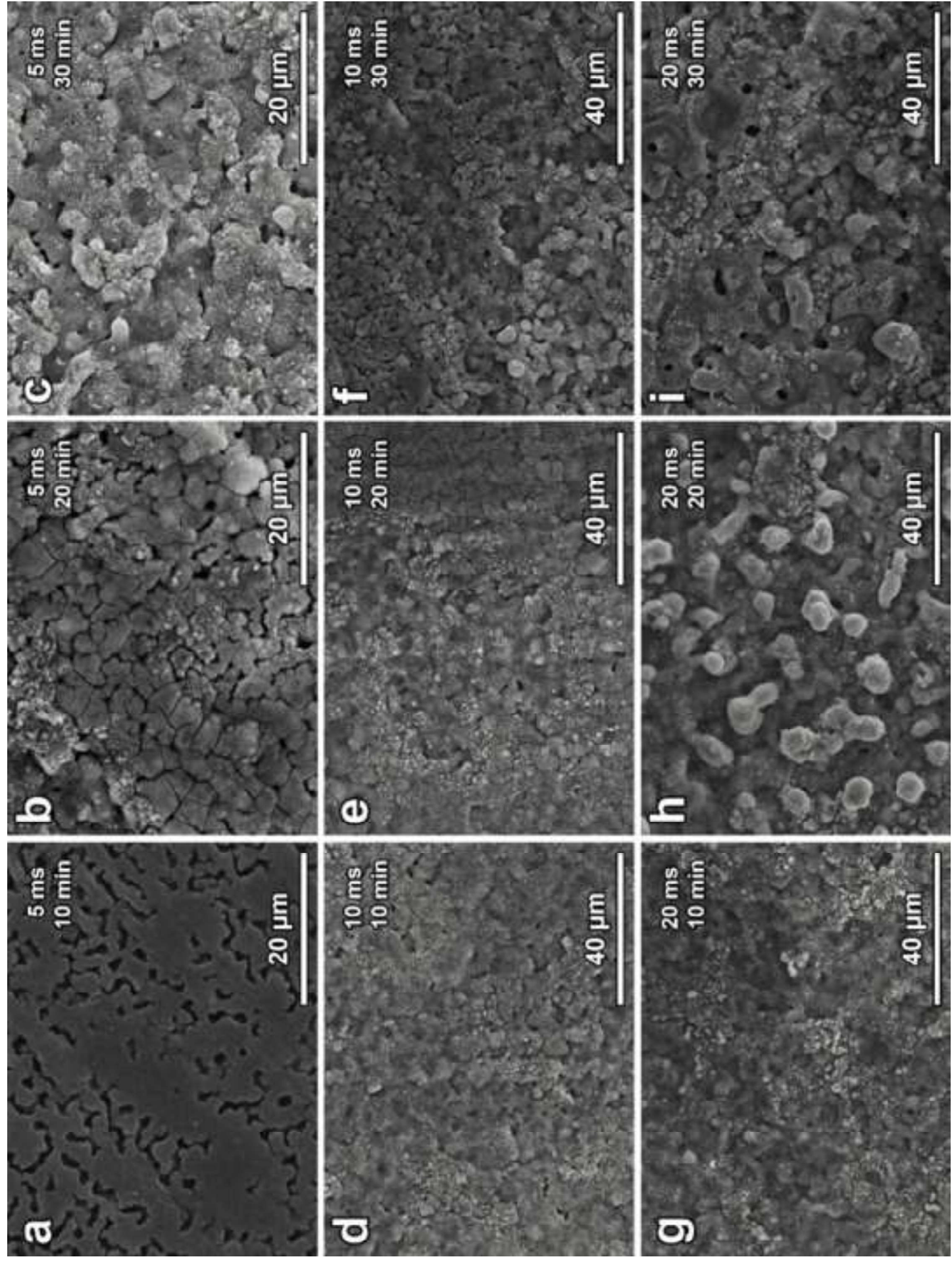

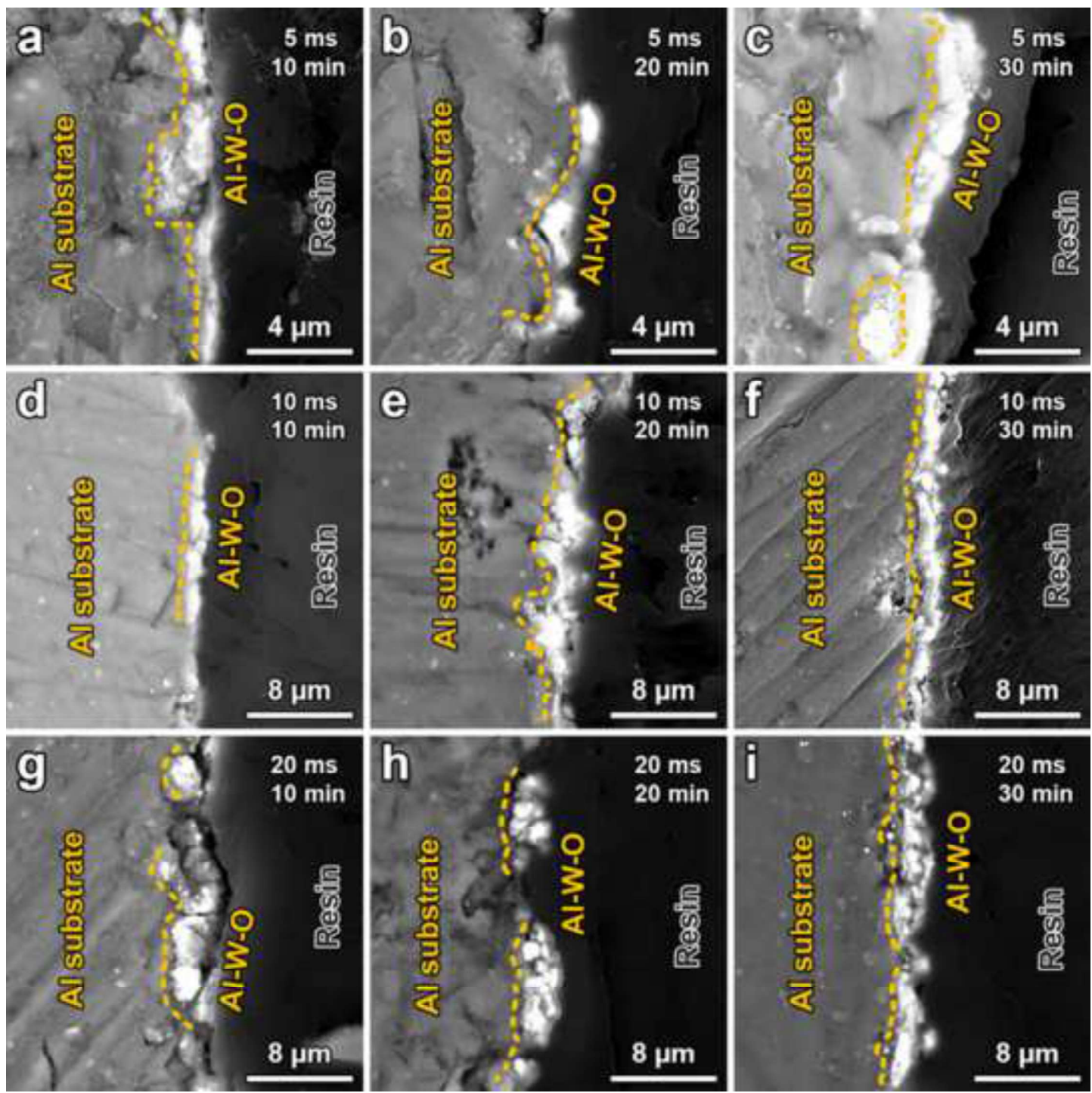

Figure 9

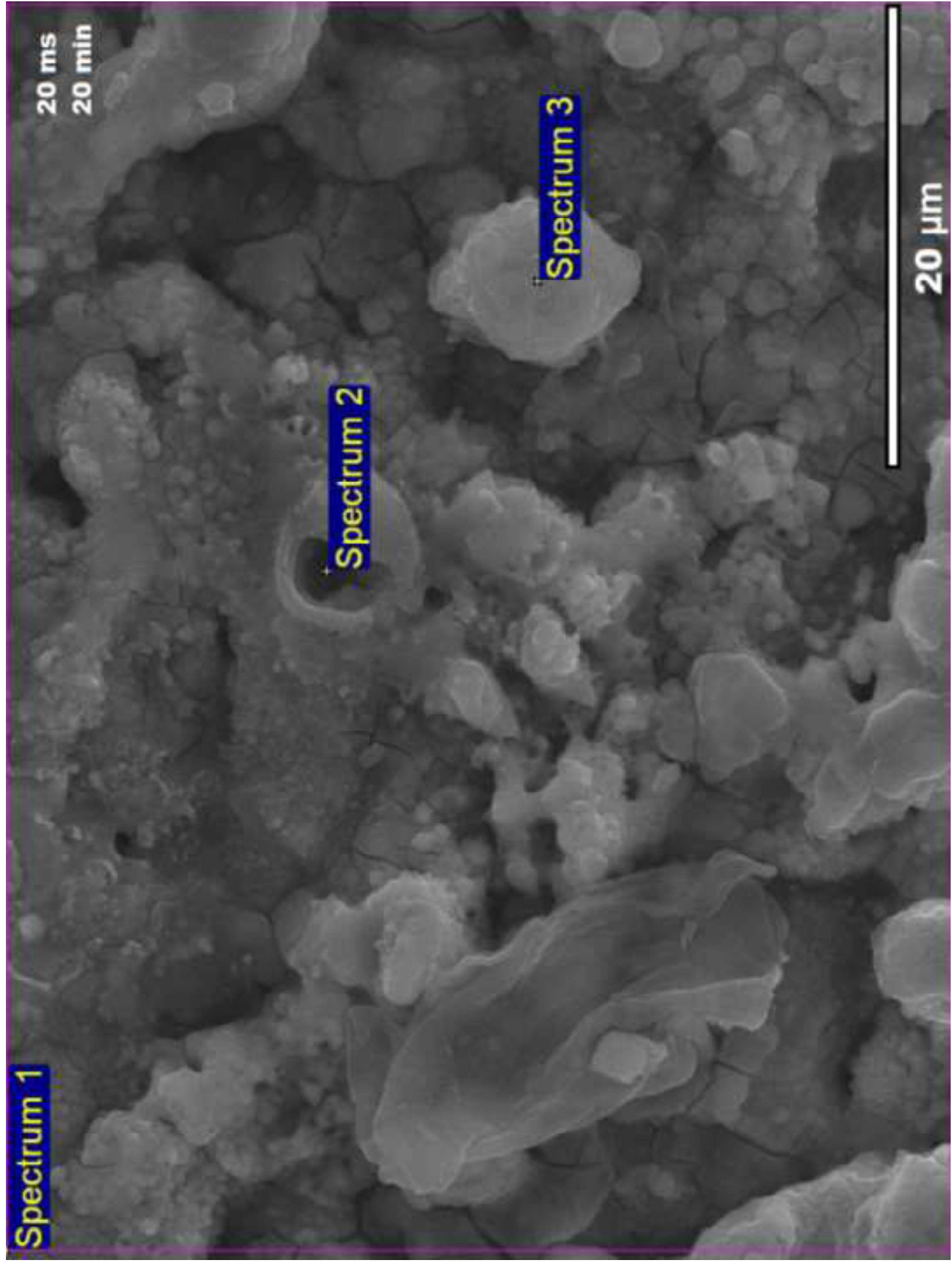

Figure 10

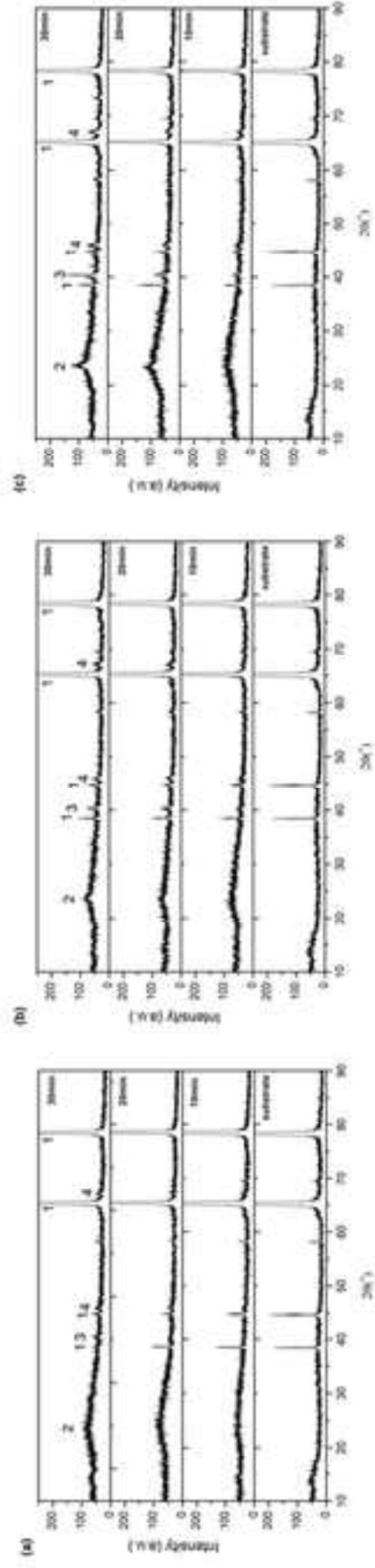

**CRediT**

Kristina Mojsilović (Investigation, Data curation, Writing – Original draft preparation), Nenad Tadić (Investigation, Data curation), Uroš Lačnjevac (Investigation, Data curation), Stevan Stojadinović (Methodology, Supervision), Rastko Vasilić (Conceptualization, Supervision, Writing – Reviewing and Editing).